 \documentclass[aps,prl,twocolumn,superscriptaddress]{revtex4}

  \usepackage{epsfig}

 \usepackage{dcolumn}
 \usepackage{bm}

\begin{document}

 \title{Spin-Orbit Coupling and Symmetry of the Order Parameter in Strontium
Ruthenate}
 \author{James F. Annett}
 \affiliation{H. H. Wills Physics Laboratory, University of Bristol,  
Tyndall Ave, BS8-1TL,
UK.}
 \author{G. Litak}
 \affiliation{Department of Mechanics, Technical University of Lublin,
Nadbystrzycka 36, PL-20-618 Lublin, Poland.}
 \author{B. L. Gy\"orffy}
 \affiliation{H. H. Wills Physics Laboratory, University of Bristol, 
Tyndall Ave, BS8-1TL,
UK.}
 \author{K. I. Wysoki\'nski}
 \affiliation{Institute of Physics, M. Curie-Sk\l{}odowska University,
Radziszewskiego 10, PL-20-031 Lublin, Poland}
 \date{ \today}

 \begin{abstract}
Determination of the orbital symmetry of a state in
spin triplet Sr$_2$RuO$_4$  superconductor 
is a challenge of considerable importance. 
Most of the experiments show that the chiral 
state of the $\hat{z} (k_x \pm ik_y)$ type is realized  
and remains stable on lowering the temperature.
Here 
we have studied the stability of various superconducting states of 
Sr$_2$RuO$_4$ in the presence of spin-orbit coupling.
 Numerically
we found that the chiral state is never the minimum energy.
Alone among the five states studied it has $\langle {\bf \hat  L}.{ \bf \hat S} \rangle=0$ and is therefore  not 
affected to linear order in the coupling parameter $\lambda$.
We found that stability of the chiral state  
requires spin dependent pairing interactions. This imposes
strong constraint on the pairing mechanism. 
\end{abstract}

\maketitle

Despite 10 years of intensive research \cite{maeno1994}, the mechanism of
 superconductivity in strontium ruthenate Sr$_2$RuO$_4$ is still not 
clarified   \cite{maeno2001,mackenzie2003}. 
A necessary first step in identifying the
effective interactions responsible for the superconductivity is to fully
identify the symmetry of the pairing state and to explain the various
relevant experimental properties below $T_c$. The current consensus is that
most experiments point towards a  p-wave spin triplet,  nodeless chiral state
of the type ${ \bf d}( { \bf k})= \hat {\bf z} (\sin k_x + {\rm i} \sin k_y)$, but with
additional lines of gap nodes somewhere on the Fermi surface \cite{mackenzie2003,eremin2004}. In
the Zhitomirsky and Rice model  \cite{zhitomirsky2001} the lines of gap
nodes arise because of an inter-band proximity effect which 
couples of the Cooper pairs between the ``active'' and ``passive'' sheets
of the Fermi surface. In our alternative proposal \cite{annett2002,annett2003}
 we introduced an effective
interlayer coupling model, which also leads to a chiral symmetry state with
lines of gap nodes. This model has a fully
gaped chiral state on the $ \gamma$ Fermi surface sheet and horizontal lines
of nodes at $k_z =  \pm  \pi/c$ on the  $ \beta$ sheet of the Fermi surface. 
Moreover, we
have shown that  it is in good agreement with the experimental
measurements  of specific heat, penetration depth and thermal conductivity 
below $T_c$.

In this letter we address the crucial question of the stability of the
chiral symmetry state in the presence of spin-orbit coupling. For p-wave
spin-triplet superfluids a large number of different pairing states are
possible, and in general many different states are degenerate in the weak
coupling limit. In the well known case of superfluid $^3$He, weak coupling
theory favors the fully gaped Balian-Werthammer (B) phase relative to the
Anderson-Brinkman-Morrel (A) phase which has point gap 
 nodes  \cite{leggett1975}. In this case the A phase is stabilized (in zero 
magnetic
field) only by a spin-fluctuation feedback effect in which the strength of the
effective pairing interaction itself is modified by the pairing symmetry.
The broken symmetry ground state effectively leads to a magnetic anisotropy
in the spin fluctuation spectrum, which in turn favors the chiral solution.
The analogous question in Sr$_2$RuO$_4$ is to ask whether the experimentally
determined chiral state is stable simply because of magnetic anisotropy
arising from spin-orbit interactions in the normal state, or whether it is
also necessary to invoke a directly spin-dependent pairing interaction.
Below we argue that this latter case applies, and therefore the very
existence of the chiral pairing state provides evidence for a magnetic
contribution to the pairing mechanism.

We follow earlier calculations 
 \cite{ng2000a,ng2000b,ng2000c,eremin2002}  where the role of spin-orbit 
coupling in the
superconducting and normal states, respectively  have been investigated.
 Our results are
 qualitatively consistent
with  the superconducting state calculations 
of Ng and Sigrist \cite{ng2000a,ng2000b,ng2000c}, but,
importantly, differ in conclusion. We show below that single-particle
 spin-orbit coupling
alone is  unlikely
 to stabilize the chiral state.
The chiral state is stabilized only after
allowing for a spin-dependent effective interaction. Such an effective
interaction is consistent with the implication of spin-orbit interaction
for transverse and longitudinal spin fluctuations, in the work of
Eremin, Manske and Bennemann \cite{eremin2002} and 
Kuwabara and Ogata \cite{kuwabara2000}. Therefore we suggest
that the experimental observation of a chiral pairing state
is conclusive proof of a spin-dependent pairing interaction.
Whilst this conclusion strongly supports the
spin fluctuation  mechanism of pairing, it could also be consistent
with other spin-dependent interaction models, 
such as Hund's rule coupling \cite{hotta2004}.

The effective pairing Hamiltonian we consider is a simple three-band,
attractive $U$, Hubbard model: 
 \begin{eqnarray}
 \hat{H}& =&  \sum_{ijmm^\prime,\sigma}  \left( ( \varepsilon_m -
 \mu) \delta_{ij} \delta_{mm^{ \prime}} - t_{mm^{ \prime}}(ij)  \right)  \hat{c}%
^+_{im \sigma} \hat{c}_{jm^{ \prime} \sigma}   \nonumber  \\
&& -  \frac{1}{2}  \sum_{ijmm^{ \prime} \sigma \sigma^{ \prime}}
U_{mm^{ \prime}}^{ \alpha \beta, \gamma \delta}(ij) 
 \hat{c}^+_{im \alpha } \hat{c}^+_{jm^ \prime \beta } \hat{c}_{jm^ \prime \gamma } 
 \hat{c}_{im \delta },
 \label{hubbard}
 \end{eqnarray}
 where $m$ and $m^{ \prime}$ refer to the three Ruthenium $t_{2g}$ orbitals 
$a=xz$, $b = yz$ and $c = xy$ and $i$ and $j$ label the sites of a body
centered tetragonal lattice. The hopping integrals $t_{mm^{ \prime}}(ij)$ and
site energies $ \varepsilon_m$ were fitted to reproduce the experimentally
determined Fermi surface  \cite{mackenzie1996,bergemann2000}.
In our previous papers  \cite{annett2002,annett2003}  we chose 
the simplest set of attractive 
pairing
interactions  $U_{mm^{ \prime}}(ij)$ which lead to
results consistent with  experiment.
However, more generally, for a spin-dependent pairing interaction the
effective Hubbard  parameters 
$U_{mm^{ \prime}}^{ \alpha \beta, \gamma \delta}(  ij)$  are spin 
as well as orbital 
dependent.

In the presence of spin-orbit coupling the original Hamiltonian (Eq. 1) should
be supplemented by the term $ \lambda { \bf \hat  L}.{ \bf \hat S}$ for each Ru atomic
site. This can be expressed \cite{ng2000a} in the form 
 \begin{equation}
H^{SO} =  \mathrm{i}  \frac{ \lambda}{2}  \sum_{i, \sigma \sigma^{ \prime}}
 \sum_{mm^{ \prime}}  \varepsilon^{ \kappa mm^{ \prime}}
 \sigma^{ \kappa}_{ \sigma \sigma^{ \prime}} c^+_{i m \sigma} c_{i
m^{ \prime} \sigma^{ \prime}},   
\label{so}
 \end{equation}
where $ \sigma^{ \kappa}_{ \sigma \sigma^{ \prime}}$, $\kappa=x,y,z$, are the  Pauli matrices, 
 $ \varepsilon^{ \kappa mm^{ \prime}}$ denotes the completely antisymmetric
tensor, and the sign convention implies that here the Ru orbital indices
must be ordered as $m=(yz,zx,xy)$ or $(b,a,c)$ in 
our notation \cite{annett2002}.

 \begin{figure}[tbp]
 \centerline{ \epsfig{file=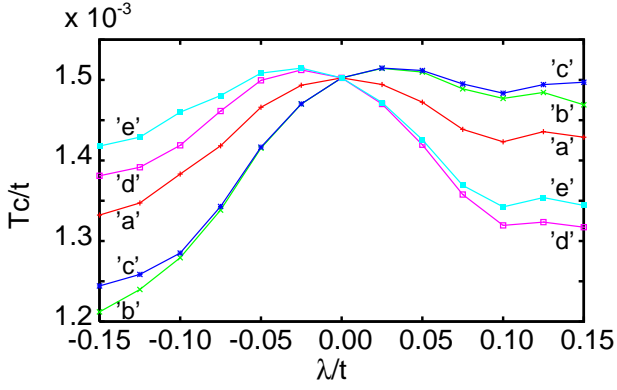,width=5.5cm,angle=-90}}
 \caption{Critical temperature for solutions of
various symmetries denoted by symbols  (a)--(e) defined in Eq. (5).
 \label{fig1}}
 \end{figure}

In the Hartree-Fock-Gorkov approximation
the gap equation takes the simplified form
 \begin{equation}
 \Delta^{ \sigma \sigma^{ \prime}}_{mm^{ \prime}}(ij) =
 U_{mm^{ \prime}}^{ \sigma \sigma^{ \prime}}(ij)
 \chi_{mm^{ \prime}}^{ \sigma \sigma^{ \prime}}(ij) \,.   
 \label{deltas}
 \end{equation}
 where the 
pairing amplitude, or order parameter, 
$ \chi_{mm^{ \prime}}^{ \sigma \sigma^{ \prime}}(ij)$, 
is defined by the usual relations  \cite{ketterson1999} 
 \begin{eqnarray}
 \chi_{mm^{ \prime}}^{ \sigma \sigma^{ \prime}}(ij) &=&  \sum_{ \nu}
u^ \nu_{im \sigma}v^{ \nu *}_{jm^{ \prime} \sigma^{ \prime}} 
(1 - 2f(E^ \nu)) \,~~~%
 \mathrm{for}~~~ \sigma= \sigma^{ \prime}   \nonumber  \\
 \chi_{mm^{ \prime}}^{ \sigma \sigma^{ \prime}}(ij)
  &=&  \frac{1}{2}  \sum_{ \nu}
(u^ \nu_{im \sigma}v^{ \nu *}_{jm^{ \prime} \sigma^{ \prime}} \label{bdg}  \\
&~& + v^{\nu *}_{im \sigma^{ \prime}}u^{ \nu }_{jm^{ \prime} \sigma}) (1 -
2f(E^ \nu)) \,~~~~~ \mathrm{for}~~~ \sigma 
 \neq  \sigma^{ \prime} .  \nonumber
 \end{eqnarray}
Note that the full spin and orbital dependent effective interaction 
$U_{mm'}^{\alpha\beta,\gamma\delta}(ij)$  
 from Eq. (\ref{hubbard}) is reduced to the simpler set of parameters
 $U_{mm^{ \prime}}^{ \sigma \sigma^{ \prime}}(ij)$
for which there are only two distinct cases: 
 $U_{mm^{ \prime}}^{ \uparrow\uparrow}(ij)=
 U_{mm^{ \prime}}^{ \downarrow\downarrow}(ij)$
 and 
$U_{mm^{ \prime}}^{ \uparrow\downarrow}(ij)=
 U_{mm^{ \prime}}^{ \downarrow\uparrow}(ij)$. 
 Other more general spin dependent terms, such as spin-flipping terms, 
 might occur, but these would not have any effect on 
 the states we consider. 

The group theory of pairing symmetry of triplet (odd-parity)
superconductivity in tetragonal crystals
has been discussed by Volovik and Gorkov \cite{volovik1985},
and is reviewed in Annett \cite{annett1990}
and Mineev and Samhokin \cite{mineev99}. 
In this case there are five distinct symmetry gap
functions which are relevant. Transforming to {\bf k}-space and
using the standard representation
$ \underline \Delta^{ \sigma \sigma^ \prime}_{ \bf k} =
 i \sigma_y \mathbf{ \sigma}.{\underline {  \bf d}}_{ \bf k}$, where $\underline \Delta$ and  $\underline {\bf d}$ are 
matrices in orbital indices, 
they have the characteristic order parameters
 \begin{eqnarray}
(a) && \hspace*{0.5cm}   {\underline {  \bf d}}_{ \bf k} \sim (0,0,X+
\mathrm{i}Y)  \nonumber  \\
 (b) && \hspace*{0.5cm}  {\underline {  \bf d}}_{ \bf k} \sim (X,Y,0)   \nonumber  \\
  (c) &&  \hspace*{0.5cm}  {\underline {  \bf d}}_{ \bf k} \sim (Y,-X,0) \label{symmetry} \\ 
 (d) &&   \hspace*{0.5cm}  {\underline {  \bf d}}_{ \bf k} \sim (X,-Y,0)  \nonumber  \\
  (e) &&  \hspace*{0.5cm}  {\underline {  \bf d}}_{ \bf k} \sim (Y,X,0)  \nonumber
 \end{eqnarray}
corresponding to the $E_{u}$, $A_{1u}$, $A_{2u}$,
 $B_{1u}$ and $B_{2u}$ pairing
symmetry states \cite{annett1990}.  Here $X({ \bf k})$
and $Y({ \bf k})$ are  pairs of basis
functions transforming as $k_x$ and $k_y$,
respectively, under crystal point group symmetries.
For in-plane interactions $X= \sin{k_x}$  and for inter-plane interactions $X= \sin{(k_x/2)}
\cos{(k_y/2)} \cos{(k_zc/2)}$
in the body centered tetragonal crystal
Sr$_2$RuO$_4$. For brevity we shall refer to
these five pairing states as (a)--(e) below.

 If the pairing interaction were spin-independent, then 
 the interaction
 parameters would obey 
 \begin{equation}
 U_{mm^{ \prime}}^{ \uparrow\uparrow}(ij)
 = U_{mm^{ \prime}}^{ \uparrow\downarrow}(ij).
 \end{equation}

\begin{figure}[tbp]
\centerline{\epsfig{file=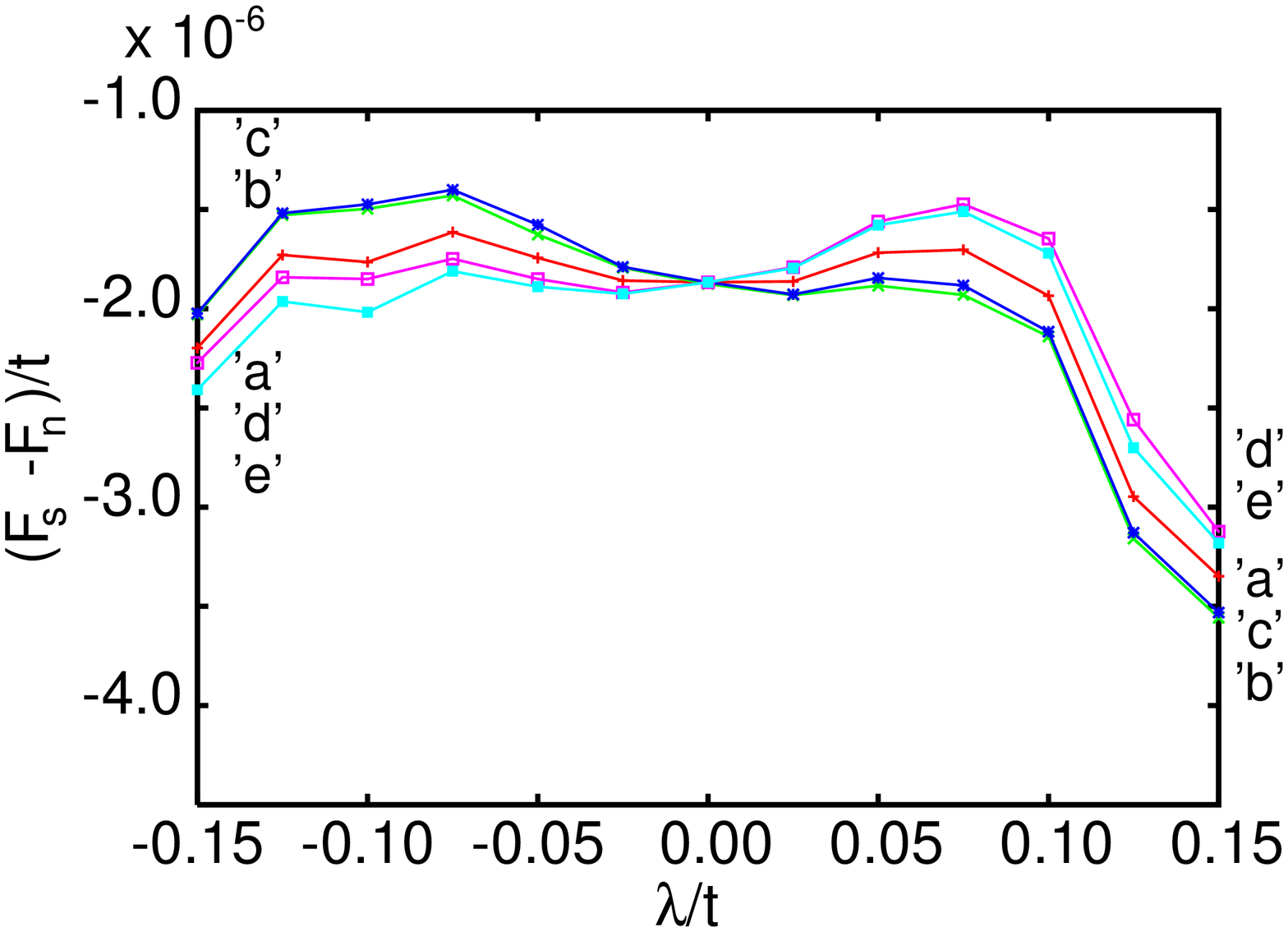,width=5.5cm,angle=-90}}

\vspace*{-3.5cm}
\hspace*{0.1cm}
\centerline{\epsfig{file=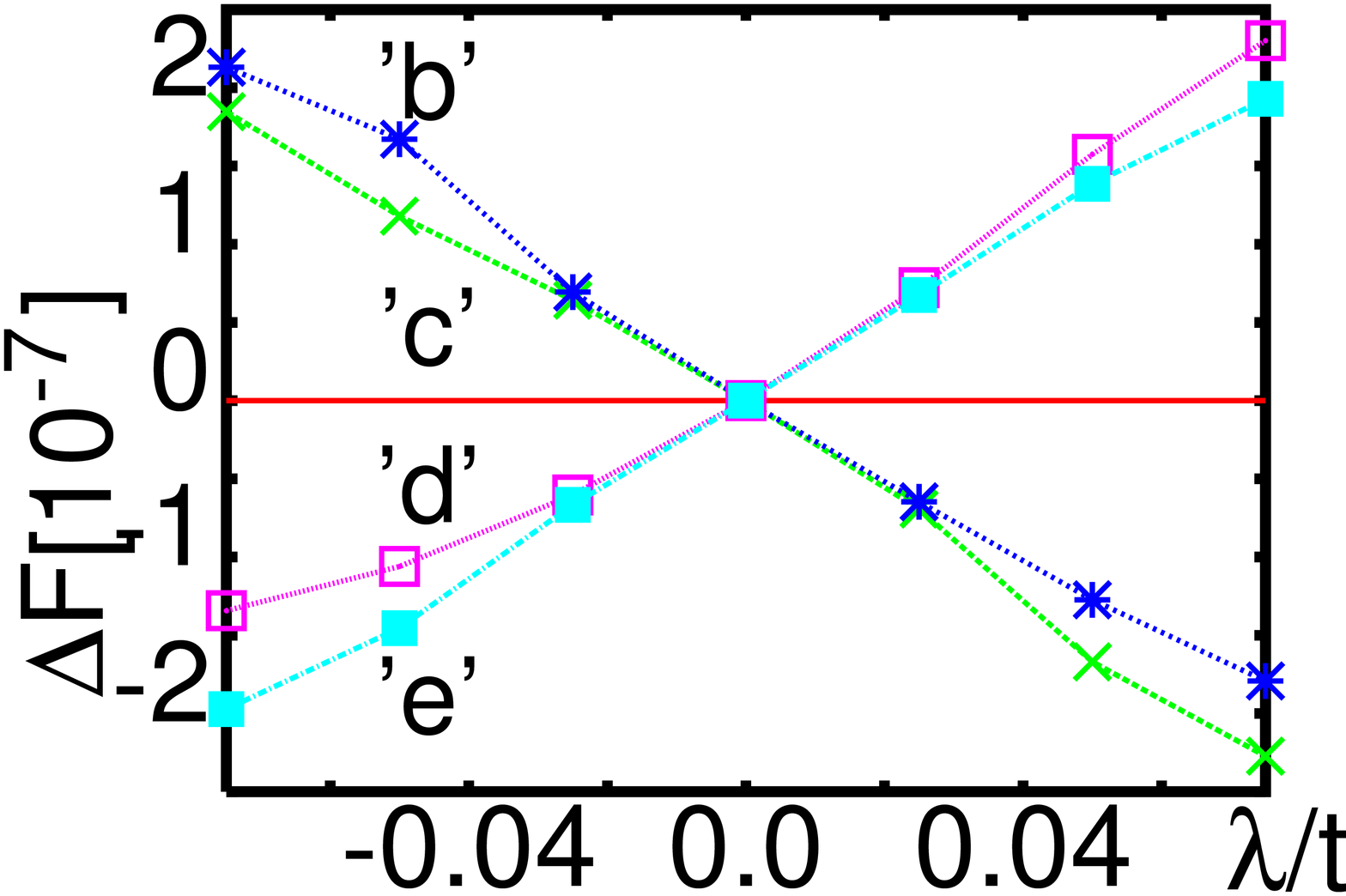,width=3.0cm,angle=-90}}
\vspace*{1.0cm}
\caption{Condensation free energy for solutions of various symmetries
defined in Eq. (\ref{symmetry}). For $\lambda > 0$ the most stable state is
(b) $\mathbf{d}(\mathbf{k})=(X,Y,0)$ , while for $\lambda < 0$
the state (e)  $\mathbf{d}(\mathbf{k})=(Y,X,0))$. The chiral state 
(a) is never stable for nonzero spin-orbit coupling. The inset shows 
the condensation energy for small $\lambda$ calculated for the states (b)--(e) with respect to
the state (a). 
\label{fig2}}
\end{figure}

\noindent
With this assumption
 we solved the full
self-consistent gap equations (Eqs. \ref{deltas}--\ref{bdg})  in presence 
of a weak spin-orbit
coupling parameter $ \lambda$. 
The band structure and the interaction parameters were chosen as in our previous works 
\cite{annett2002,annett2003,wysokinski2003}.
The results for the critical temperature $T_c$
are shown in Fig. \ref{fig1} and the condensation free energies ($F_s-F_n$)
at $T=0$ are shown in Fig. \ref{fig2}. Both figures show clearly that all
five pairing states become exactly degenerate at $\lambda=0$, as expected.
For positive values of $\lambda$ the states split in such a way that the (b)
and (c) states have the highest $T_c$ and the greatest (negative)
condensation energy. Conversely, for negative values of $ \lambda$ states (d)
and (e) have the highest $T_c$ and the greatest condensation energy.
Surprisingly, the chiral state, (a), is  \emph{never} stable compared to the
others. 
 
This counter-intuitive result deserves some comment. Firstly, one should
note that experiments point unequivocally to a chiral solution, since only
this one would have the constant in-plane spin susceptibility as measured
in both NMR \cite{ishida2001} and neutron scattering \cite{duffy2000} 
experiments.
Secondly, Ng and Sigrist \cite{ng2000a,ng2000b,ng2000c} have argued on the
basis of a very similar model to ours that 
spin-orbit interaction  \emph{does}
stabilizes the chiral phase. 

Although we draw different conclusions, 
on closer inspection, our results are
indeed very similar to theirs. They used a three-band model characterized by
two main pairing parameters $V_1=V_2$ and $V_3$. They found that the chiral
state is favored for positive $ \lambda$ only in the region $V_3/V_2 < 1$.
For $V_3/V_2 > 1$ they found that states (b) and (c) (III and II in their
notation) were favored, and that these two states were nearly degenerate
being separated only by an additional inter-band proximity effect coupling
term. Our results are qualitatively consistent with this finding, if we
interpret our effective Hubbard 
interaction terms \cite{annett2002,annett2003} $U_ \perp$ and $U_ \parallel$
as giving a corresponding value 
of $V_3/V_2  \approx 2 U_\perp/U_\parallel =2.4$. 
In our model we have no inter-band proximity effect coupling, and hence
states (b) and (c) are essentially degenerate.

The physical values of the spin-orbit coupling parameter are positive, and
believed to be of order $ \lambda =40-80$meV, corresponding to
$ \lambda/t \simeq 0.5-1$ in our units.
We have also carried out fully self-consistent
calculations at $ \lambda =0.5$, and found the same
ordering of the ground
states as shown in Fig. \ref{fig2}. But for such large values
of $ \lambda $ we
observed that the Fermi surface shape begins to change, and so it was
necessary to refit the Hubbard model hopping parameters to the experimental
Fermi surface 
\cite{bergemann2000}. However this refitting of the Fermi surface did not
change
the results qualitatively from those expected from Fig. \ref{fig2},
and so we   
conclude that the precise Fermi surface shape is not essential in
determining the relative stability of the different pairing states.

In order to understand more fully the physical origin of the spin-orbit splitting shown 
in Figs. \ref{fig1}--\ref{fig2} we consider small perturbations about the point $\lambda=0$. Here the change in condensation 
energy to the linear order is
\begin{eqnarray}
\delta F = \frac{\lambda}{2} \langle {\bf  \hat  L}.{\bf  \hat S} \rangle =  \frac{\lambda}{2} \sum_{m,m'}  
\langle m| \hat L_z|m' \rangle 
n_{mm'}^{\uparrow \uparrow} \nonumber \\ 
+   \frac{\lambda}{2} \sum_{m,m'}  \langle m| \hat L_x-{\rm i} \hat L_y  |m' \rangle n_{mm'}^{\uparrow
\downarrow} \\
+   \frac{\lambda}{2} \sum_{m,m'}  \langle m| \hat L_x+{\rm i} \hat L_y  |m' \rangle n_{mm'}^{\downarrow \uparrow} \nonumber 
\\-   \frac{\lambda}{2} \sum_{m,m'} \langle m| \hat L_z  |m' \rangle n_{mm'}^{\downarrow \downarrow}, \nonumber 
\end{eqnarray}
where $n^{\sigma,\sigma'}_{mm'} = \langle  c_{im\sigma}^+ c_{im'\sigma'} \rangle $ are on-site density matrices for the Ru $d$ 
orbitals, 
in 
the  
absence of spin-orbit coupling.
In fact, in our model, the $\gamma$ band associated with the $d_{xy}$ orbital is strongly decoupled from the $d_{xz}$ and  
$d_{yz}$
 orbitals, and  so the only significant off-diagonal density matrix elements are $n_{ab}^{\sigma\sigma'}$ and 
$n_{ba}^{\sigma\sigma'}$.
Thus we find the spin-orbit correction
\begin{equation}
\delta F= \frac{\lambda}{2} {\rm i} \left( n_{ab}^{\uparrow \uparrow} -n_{ab}^{\downarrow \downarrow}-  n_{ba}^{\uparrow 
\uparrow} + n_{ba}^{\downarrow \downarrow} \right). 
\end{equation} For the chiral (a) state  $n_{ab}^{\uparrow \uparrow}= n_{ab}^{\downarrow \downarrow}$ and so
the spin-orbit coupling is zero to linear order in $\lambda$. On the other hand for (b)--(e) 
$n_{ab}^{\uparrow \uparrow}= -n_{ab}^{\downarrow \downarrow}$ and so the spin-orbit energy is of order $\lambda$. 
This explains the small $\lambda$ behavior 
seen in our numerical calculations, and shown in the inset to Fig. \ref{fig2}. It is interesting to note that for all the 
states (a)--(e)
the expectation value of the spin $\langle  { \bf \hat S}  \rangle=0$ for each orbital $m$. On the other hand the z component 
of the 
orbital 
angular momentum is
\begin{equation}
\langle \hat L_z \rangle =  n_{ab}^{\uparrow \uparrow} +n_{ab}^{\downarrow \downarrow}-   \left(n_{ba}^{\uparrow
\uparrow} + n_{ba}^{\downarrow \downarrow}  \right),
\end{equation}  
which is non-zero only for the chiral state. Therefore $ \langle {\bf   \hat L}.{\bf  \hat S} \rangle$ is non-zero for four 
states 
(b)--(e) which 
have $\langle {\bf   \hat S } \rangle=\langle {\bf \hat L} \rangle=0$ while the (a) state has $\langle {\bf \hat  L}.{\bf  
\hat 
S} \rangle=0$, 
$\langle  
\hat L_z \rangle 
\neq 0$.

Given this surprising result, how we can understand the stability of the chiral state, which is 
required for understanding the susceptibility experiments?
 Clearly, if the effective pairing interaction
 is caused, even in part, by spin fluctuations, then in general we
 would expect
 that
$U_{mm^{ \prime}}^{ \uparrow\uparrow}(ij) 
 \neq U_{mm^{ \prime}}^{ \uparrow\downarrow}(ij)$, corresponding to the two Fermi liquid parameters  $F^{\uparrow \uparrow}$, 
  $F^{\uparrow \downarrow}$ \cite{leggett1975}.
In the spin-fluctuation theory of Kuwabara and Ogata \cite{kuwabara2000}
one can see that the parallel spin and opposite spin
effective interactions are derived from
transverse or longitudinal spin fluctuations, respectively, and hence
they become different when spin-orbit interactions are present.
The role of spin-orbit coupling in the transverse,
$\chi^ \pm({ \bf q})$, and longitudinal, $ \chi^z({ \bf q})$,   
spin susceptibilities has been explored more fully by   
Eremin, Manske and Bennemann \cite{eremin2002} and their theory agrees well
with the results of NMR \cite{ishida2001} and
neutron scattering experiments \cite{braden2003}.
  Therefore spin-orbit coupling has two
 quite distinct effects on a spin triplet superconductor.
Firstly, the spin-orbit term, Eq. (\ref{so}), directly enters
the gap equation leading to
small changes in the single-particle band structure and density of states,
and secondly, it leads to spin anisotropy in the effective pairing interaction.
 Both of these effects must be included in order to establish the relative
 stability of the different possible pairing symmetry states \cite{yanase2003}.

\begin{figure}[tbp]
 \centerline{ \epsfig{file=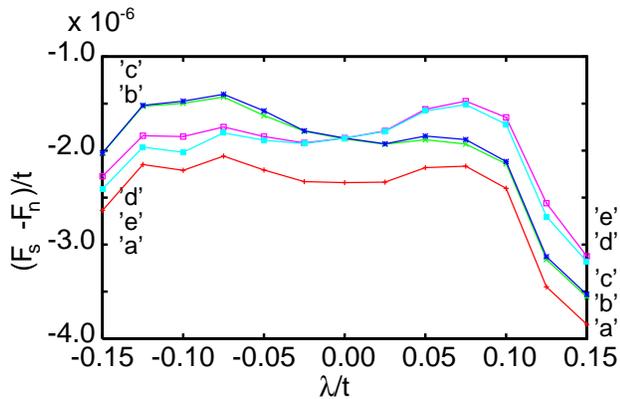,width=5.5cm,angle=-90}} 
 \caption{The same as in Fig. \ref{fig2}, but here we 
 assumed that interactions for 'a' solution $
U^{ \prime}$ are slightly (by 1\%) larger than  for other solutions $U$: 
$U^{ \prime}/U=1.01$. 
 \label{fig3}}
\vspace*{-0.5cm}
 \end{figure}

\begin{figure}[tbp]
 \centerline{ \epsfig{file=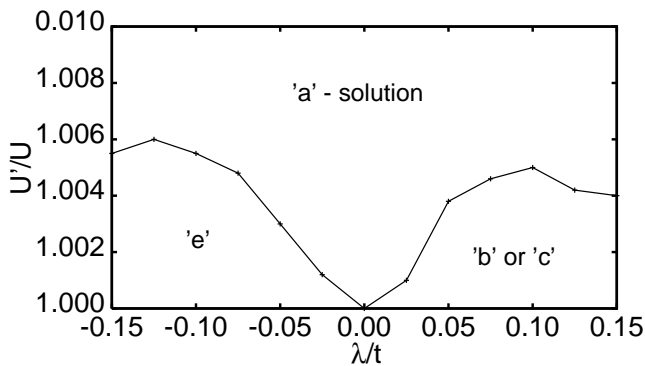,width=5.5cm,angle=-90}} 
 \caption{Phase diagram for different interaction ratio $U^{ \prime}/U$ and
a spin-orbit coupling $ \protect \lambda$. 
 \label{fig4}}
 \end{figure}

 We have found that 
the chiral state becomes stable when we increased the interaction 
 $U^{\uparrow\downarrow}=U'$ by 1\% with 
 respect to $U^{ \uparrow \uparrow}=U$. These results  in Fig. \ref{fig3}. 
The phase diagram in the $U'/U$ vs. $\lambda/t $  space
is presented in Fig. \ref{fig4}. It shows that to stabilize
the chiral state in the presence of sizeable 
spin-orbit coupling requires increase of $U'$ by less than 1\%.

It follows from  these results, that the pairing interaction
itself must directly favor the chiral symmetry state
and
would be an analogue of the well known spin-fluctuation feedback mechanism
which stabilizes the A-phase in superfluid $^{3}$He \cite{leggett1975}. Eremin, Manske and
Bennemann \cite{eremin2002} calculated the spin susceptibility in the normal
state of Sr$_{2}$RuO$_{4}$. They found that spin-orbit interaction led to a
magnetic anisotropy between 
$ \chi ^{z}( \mathbf{q}, \omega )$ and $ \chi ^{ \pm}( \mathbf{q}, \omega )$. 
The well known argument for $^{3}$He would then
suggest that these susceptibilities lead to different effective pairing
interactions between spin-triplet, $S=1$, quasiparticle pairs 
in  the $m=0$ or $m= \pm 1$ channels. As it turns out a small  increase
of the interaction, $U'/U=1.01$, stabilizes the (a) state.

This work has been partially supported by the KBN  grant No. 2P03B06225 and the NATO
Collaborative Linkage Grant No. 979446 and the INTAS grant No. 01-654.
GL would like to thank  Max Planck Institute for the Physics of Complex 
Systems in
Dresden for hospitality. The authors are grateful D. Manske and I. Eremin for 
discussions.

 \end{document}